\documentclass[floatfix,10pt,reprint,amsmath,amssymb,superscriptaddress,aps,prd,nofootinbib]{revtex4-2}
\usepackage{graphicx}
\usepackage{dcolumn}
\usepackage{bm}
\usepackage[usenames]{color}
\usepackage{lineno}
\usepackage{etoolbox} 
\usepackage{url} 
\usepackage{placeins}
\usepackage{hyperref} 
\newcommand*{\Apr}{{A^\prime}}
\newcommand*{\Zpr}{{Z^\prime}}

\newcommand*\linenomathpatch[1]{%
  \cspreto{#1}{\linenomath}%
  \cspreto{#1*}{\linenomath}%
  \csappto{end#1}{\endlinenomath}%
  \csappto{end#1*}{\endlinenomath}%
}
\linenomathpatch{equation}
\linenomathpatch{gather}
\linenomathpatch{multline}
\linenomathpatch{align}
\linenomathpatch{alignat}
\linenomathpatch{flalign}

\def\address{\@ifstar{\address@star}%
  {\@ifnextchar[{\address@optarg}{\address@noptarg}}}

\begin{document}


\title{Search for a light $Z^\prime$ in the $L_\mu-L_\tau$ scenario with the   NA64-$e$ experiment at CERN} 

\author{Yu.~M.~Andreev}\affiliation{ Institute for Nuclear Research, 117312 Moscow, Russia}
\author{D.~Banerjee}\affiliation{ CERN, European Organization for Nuclear Research, CH-1211 Geneva, Switzerland}
\author{B.~Banto Oberhauser}\affiliation{ CERN, European Organization for Nuclear Research, CH-1211 Geneva, Switzerland}
\author{J.~Bernhard}\affiliation{ ETH Z\"urich, Institute for Particle Physics and Astrophysics, CH-8093 Z\"urich, Switzerland}
\author{P.~Bisio}\affiliation{INFN, Sezione di Genova, 16147 Genova, Italia}\affiliation{Universit\'a degli Studi di Genova, 16126 Genova, Italy}
\author{M.~Bond\'i}\affiliation{INFN, Sezione di Roma Tor Vergata, 00133 Rome, Italy}\affiliation{Universit\'a di Roma Tor Vergata, 00133 Rome Italy}
\author{V.~E.~Burtsev}\affiliation{ Joint Institute for Nuclear Research, 141980 Dubna, Russia}
\author{A.~Celentano}\affiliation{INFN, Sezione di Genova, 16147 Genova, Italia}
\author{N.~Charitonidis}\affiliation{ CERN, European Organization for Nuclear Research, CH-1211 Geneva, Switzerland}
\author{A.~G.~Chumakov}\affiliation{ Tomsk Polytechnic University, 634050 Tomsk, Russia}\affiliation{ Tomsk State Pedagogical University, 634061 Tomsk, Russia}
\author{D.~Cooke}\affiliation{ UCL Departement of Physics and Astronomy, University College London, Gower St. London WC1E 6BT, United Kingdom}
\author{P.~Crivelli}\affiliation{ ETH Z\"urich, Institute for Particle Physics and Astrophysics, CH-8093 Z\"urich, Switzerland}
\author{E.~Depero}\affiliation{ ETH Z\"urich, Institute for Particle Physics and Astrophysics, CH-8093 Z\"urich, Switzerland}
\author{A.~V.~Dermenev}\affiliation{ Institute for Nuclear Research, 117312 Moscow, Russia}
\author{S.~V.~Donskov}\affiliation{ State Scientific Center of the Russian Federation Institute for High Energy Physics of National Research Center 'Kurchatov Institute' (IHEP), 142281 Protvino, Russia}
\author{R.~R.~Dusaev}\affiliation{ Tomsk Polytechnic University, 634050 Tomsk, Russia}
\author{T.~Enik}\affiliation{  Joint Institute for Nuclear Research, 141980 Dubna, Russia}
\author{V.~N.~Frolov}\affiliation{  Joint Institute for Nuclear Research, 141980 Dubna, Russia}
\author{A.~Gardikiotis}\affiliation{ Physics Department, University of Patras, 265 04 Patras, Greece}
\author{S.~G.~Gerassimov }\affiliation{ Technische Universit\"at M\"unchen, Physik  Department, 85748 Garching, Germany}\affiliation{ P.N.Lebedev Physical Institute of the Russian Academy of Sciences, 119 991 Moscow, Russia}
\author{S.~N.~Gninenko}\affiliation{ Institute for Nuclear Research, 117312 Moscow, Russia}
\author{M.~H\"osgen}\affiliation{ Universit\"at Bonn, Helmholtz-Institut f\"ur Strahlen-und Kernphysik, 53115 Bonn, Germany}
\author{M.~Jeckel}\affiliation{ CERN, European Organization for Nuclear Research, CH-1211 Geneva, Switzerland}
\author{V.~A.~Kachanov}\affiliation{ State Scientific Center of the Russian Federation Institute for High Energy Physics of National Research Center 'Kurchatov Institute' (IHEP), 142281 Protvino, Russia}
\author{A.~E.~Karneyeu}\affiliation{ Institute for Nuclear Research, 117312 Moscow, Russia}
\author{G.~Kekelidze}\affiliation{  Joint Institute for Nuclear Research, 141980 Dubna, Russia}
\author{B.~Ketzer}\affiliation{ Universit\"at Bonn, Helmholtz-Institut f\"ur Strahlen-und Kernphysik, 53115 Bonn, Germany}
\author{D.~V.~Kirpichnikov}\affiliation{ Institute for Nuclear Research, 117312 Moscow, Russia}
\author{M.~M.~Kirsanov}\affiliation{ Institute for Nuclear Research, 117312 Moscow, Russia}
\author{V.~N.~Kolosov}\affiliation{ State Scientific Center of the Russian Federation Institute for High Energy Physics of National Research Center 'Kurchatov Institute' (IHEP), 142281 Protvino, Russia}
\author{I.~V.~Konorov}\affiliation{ Technische Universit\"at M\"unchen, Physik  Department, 85748 Garching, Germany}\affiliation{ P.N.Lebedev Physical Institute of the Russian Academy of Sciences, 119 991 Moscow, Russia}
\author{S.~G.~Kovalenko}\affiliation{ Departamento de Ciencias F\'{i}sicas, Universidad Andres Bello, Sazi\'{e} 2212, Piso 7, Santiago, Chile}\affiliation{ Millennium Institute for Subatomic Physics at the High-Energy Frontier (SAPHIR),  ICN2019\_044, ANID, Chile}
\author{V.~A.~Kramarenko}\affiliation{  Joint Institute for Nuclear Research, 141980 Dubna, Russia}\affiliation{ Skobeltsyn Institute of Nuclear Physics, Lomonosov Moscow State University, 119991  Moscow, Russia}
\author{L.~V.~Kravchuk}\affiliation{ Institute for Nuclear Research, 117312 Moscow, Russia}
\author{ N.~V.~Krasnikov}\affiliation{  Joint Institute for Nuclear Research, 141980 Dubna, Russia}\affiliation{ Institute for Nuclear Research, 117312 Moscow, Russia}
\author{S.~V.~Kuleshov}\affiliation{ Departamento de Ciencias F\'{i}sicas, Universidad Andres Bello, Sazi\'{e} 2212, Piso 7, Santiago, Chile}\affiliation{ Millennium Institute for Subatomic Physics at the High-Energy Frontier (SAPHIR),  ICN2019\_044, ANID, Chile}
\author{V.~E.~Lyubovitskij}\affiliation{ Tomsk Polytechnic University, 634050 Tomsk, Russia}\affiliation{ Tomsk State Pedagogical University, 634061 Tomsk, Russia}\affiliation{ Universidad T\'{e}cnica Federico Santa Mar\'{i}a and CCTVal, 2390123 Valpara\'{i}so, Chile}\affiliation{ Millennium Institute for Subatomic Physics at the High-Energy Frontier (SAPHIR),  ICN2019\_044, ANID, Chile}
\author{V.~Lysan}\affiliation{  Joint Institute for Nuclear Research, 141980 Dubna, Russia} 
\author{L.~Marsicano}\thanks{Corresponding author}\email{luca.marsicano@ge.infn.it}\affiliation{INFN, Sezione di Genova, 16147 Genova, Italia} 
\author{V.~A.~Matveev}\affiliation{  Joint Institute for Nuclear Research, 141980 Dubna, Russia}
\author{Yu.~V.~Mikhailov}\affiliation{ State Scientific Center of the Russian Federation Institute for High Energy Physics of National Research Center 'Kurchatov Institute' (IHEP), 142281 Protvino, Russia}
\author{L.~Molina Bueno}\affiliation{ ETH Z\"urich, Institute for Particle Physics and Astrophysics, CH-8093 Z\"urich, Switzerland}\affiliation{Instituto de Fisica Corpuscular (CSIC/UV), Carrer del Catedrátic José Beltrán Martinez, 2, 46980 Paterna, Valencia}
\author{D.~V.~Peshekhonov}\affiliation{  Joint Institute for Nuclear Research, 141980 Dubna, Russia}
\author{V.~A.~Polyakov}\affiliation{ State Scientific Center of the Russian Federation Institute for High Energy Physics of National Research Center 'Kurchatov Institute' (IHEP), 142281 Protvino, Russia}
\author{B.~Radics}\affiliation{ ETH Z\"urich, Institute for Particle Physics and Astrophysics, CH-8093 Z\"urich, Switzerland}
\author{R.~Rojas}\affiliation{ Universidad T\'{e}cnica Federico Santa Mar\'{i}a and CCTVal, 2390123 Valpara\'{i}so, Chile}
\author{A.~Rubbia}\affiliation{ ETH Z\"urich, Institute for Particle Physics and Astrophysics, CH-8093 Z\"urich, Switzerland}
\author{K.~M.~Salamatin}\affiliation{ Joint Institute for Nuclear Research, 141980 Dubna, Russia}
\author{V.~D.~Samoylenko}\affiliation{ State Scientific Center of the Russian Federation Institute for High Energy Physics of National Research Center 'Kurchatov Institute' (IHEP), 142281 Protvino, Russia}
\author{H.~Sieber}\affiliation{ ETH Z\"urich, Institute for Particle Physics and Astrophysics, CH-8093 Z\"urich, Switzerland}
\author{D.~Shchukin}\affiliation{ P.N.Lebedev Physical Institute of the Russian Academy of Sciences, 119 991 Moscow, Russia}
\author{V.~O.~Tikhomirov}\affiliation{ P.N.Lebedev Physical Institute of the Russian Academy of Sciences, 119 991 Moscow, Russia}
\author{I.~Tlisova}\affiliation{ Institute for Nuclear Research, 117312 Moscow, Russia} 
\author{A.~N.~Toropin}\affiliation{ Institute for Nuclear Research, 117312 Moscow, Russia}
\author{A.~Yu.~Trifonov}\affiliation{ Tomsk Polytechnic University, 634050 Tomsk, Russia}\affiliation{ Tomsk State Pedagogical University, 634061 Tomsk, Russia}
\author{P.~Ulloa}\affiliation{ Departamento de Ciencias F\'{i}sicas, Universidad Andres Bello, Sazi\'{e} 2212, Piso 7, Santiago, Chile}
\author{B.~I.~Vasilishin}\affiliation{ Tomsk Polytechnic  University, 634050 Tomsk, Russia}
\author{G.~Vasquez Arenas}\affiliation{ Universidad T\'{e}cnica Federico Santa Mar\'{i}a and CCTVal, 2390123 Valpara\'{i}so, Chile}
\author{P.~V.~Volkov}\affiliation{  Joint Institute for Nuclear Research, 141980 Dubna, Russia}\affiliation{ Skobeltsyn Institute of Nuclear Physics, Lomonosov Moscow State University, 119991  Moscow, Russia}
\author{V.~Yu.~Volkov}\affiliation{ Skobeltsyn Institute of Nuclear Physics, Lomonosov Moscow State University, 119991  Moscow, Russia}
\author{I.~Voronchikhin}\affiliation{ Tomsk Polytechnic University, 634050 Tomsk, Russia}\affiliation{ Tomsk State Pedagogical University, 634061 Tomsk, Russia}
%

\collaboration{The NA64 collaboration}
\date{\today}
\begin{abstract}

The extension of Standard Model made by inclusion of additional $U(1)$ gauge $L_\mu-L_\tau$ symmetry can explain the difference between the measured and the predicted value of the muon magnetic moment and solve the tension in $B$ meson decays. This model predicts the existence of a new, light $Z^\prime$ vector boson, predominantly coupled to second and third generation leptons, whose interaction with electrons is due to a loop mechanism involving muons and taus. 
In this work, we present a rigorous evaluation of the upper limits in the $\Zpr$ parameter space, obtained from the analysis of the data collected by the NA64-$e$ experiment at CERN SPS, that performed a search for light dark matter with $2.84\times10^{11}$ electrons impinging with 100 GeV on an active thick target. 
The resulting limits 
touch the muon $g-2$ preferred band for values of the $\Zpr$ mass of order of 1 MeV, while the sensitivity projections for the future high-statistics NA64-$e$ runs demonstrate the power of the electrons/positron beam approach in this theoretical scenario. 
\end{abstract}

\maketitle

\section{\label{sec:intro} Introduction}
The Standard Model (SM) of particle physics works remarkably well in describing and interpreting the experimental results provided by different, complementary efforts, operating at different energy scales~\cite{Butterworth:2016mrp}. However, recent years have been marked by 
potential experimental signs of new physics phenomena, the so-called ``anomalies'', which cannot be explained within the SM, calling for the development of extensions beyond SM providing a more accurate description of Nature.

Among these, a remarkable example is provided by the recent measurement of the muon magnetic moment $a_\mu \equiv (g_\mu-2)/2$ reported by the Fermilab E989 experiment~\cite{Muong-2:2021ojo}, that, combined with the original BNL result~\cite{Muong-2:2006rrc}, leads to a $4.2\sigma$ discrepancy with the most-updated theoretical prediction computed by the Muon $g-2$ Theory Initiative~\cite{Aoyama:2020ynm}, $a_\mu \mathrm{(Exp)}-a_\mu \mathrm{(SM)} = (251 \pm 59)\times 10^{-11}$. Even if recent alternative results obtained through lattice QCD calculations may possibly release this tension~\cite{Borsanyi_2021}, still the muon magnetic moment anomaly motivates the interest toward beyond SM scenarios that may explain it.
In this work, we consider the SM extension in which the anomaly-free combination $L_\mu-L_\tau$ is associated to a new $U(1)$ gauge symmetry, thus introducing a new massive vector boson $Z^\prime$ coupled to the difference between the second and third generation leptonic currents~\cite{He:1990pn,He:1991qd}. 
The corresponding new lagrangian terms read~\cite{Bauer:2018onh}:
\begin{align}
    \mathcal{L} \subset & -\frac{1}{4}\Zpr_{\mu\nu}{\Zpr}^{\mu\nu} +\frac{1}{2}m^2_\Zpr \Zpr_\mu \Zpr^\mu + \\ \nonumber
    & -  g_\Zpr \Zpr_\mu \left (\overline{\mu}\gamma^\mu\mu +\overline{\nu}_\mu\gamma^\mu P_L \nu_\mu  
    -\overline{\tau}\gamma^\mu\tau
    -\overline{\nu}_\tau\gamma^\mu P_L \nu_\tau  
    \right) \;\;,
\end{align}
where $Z^\prime_{\mu\nu}\equiv\partial_\mu \Zpr_\nu-\partial_\nu \Zpr_\mu$ is the $\Zpr$ field strength, $m_\Zpr$ is the $\Zpr$ mass, $P_L=(1-\gamma_5)/2$, and $g_\Zpr$ is the coupling between the $\Zpr$ boson and the $L_\mu-L_\tau$ SM current. 

At leading order, the $\Zpr$ contributes to the muon magnetic moment as:
\begin{equation}
    \delta a_\mu = \frac{g^2_\Zpr}{8\pi^2}F\left(m_\Zpr/m_\mu\right) \; \;,
\end{equation}
where $F(x)=\int_0^1\,dz\frac{2z(1-z)^2}{(1-z)^2+x^2z}$ \cite{Gninenko:2001hx,Patra:2016shz,Biswas:2016yjr}. 
This can explain the observed muon magnetic moment discrepancy if the parameter $m_\Zpr$ and $g_\Zpr$ lie in a well defined area of the parameters space, roughly defined by $g_\Zpr \in [3.2,5.5]\times 10^{-4}$ at $2 \sigma$ for $m_\Zpr \ll m_\mu$~\cite{Amaral:2021rzw}. The lack of a tree-level coupling with the SM electron also means that the model does not contribute appreciably to the $e^-$ magnetic moment, in agreement with the experimental observation~\cite{CarcamoHernandez:2019ydc}. The $\Zpr$ model, either in the ``vanilla'' form described before or in association with more elaborated SM extensions, has also been advocated to explain other SM anomalies, such as the $B$ decay anomaly~\cite{Crivellin:2015mga,Altmannshofer:2016jzy,Ko:2017yrd,Baek:2017sew} and the lepton-flavor universality violation~\cite{Heeck:2022znj,Greljo:2021xmg,Greljo:2021npi}. The $\Zpr$ model  has also been connected to the Dark Matter (DM)~\cite{Patra:2016shz,Biswas:2016yan,Foldenauer:2018zrz,Arcadi:2018tly,Kamada:2018zxi,Kahn:2018cqs,Holst:2021lzm} and to the neutrino mass phenomenology~\cite{Baek:2015mna}.

These arguments recently motivated a large number of complementary efforts to search for the $\Zpr$, either by performing a re-analysis of existing experimental datasets, or proposing new, dedicated experiments. The BaBar~
\cite{BaBar:2016sci} and CMS~\cite{CMS:2018yxg} experiments investigated existence of the $\Zpr$ by exploiting the visible decay channel $\Zpr\rightarrow \mu^+\mu^-$, searching for a resonance peak in the dimuon mass distribution, on top of the SM background. The Belle-II experiment focused instead on the $\Zpr$ invisible decay channel, exploiting the reaction $e^+e^-\rightarrow \mu^+\mu^-\Zpr$, where the $\Zpr$ is radiated by one of the final state muons, searching for a resonance peak in the recoil mass of the final state muons~\cite{Belle-II:2019qfb}. Dedicated $\Zpr$ searches at Belle-II exploiting mono-photon signatures have also been suggested~\cite{Kaneta:2016uyt,Banerjee:2018mnw,Campajola:2021pdl}.  Stringent upper limits on the $g_\Zpr$ coupling have also been obtained by neutrino experiments, such as CCFR~\cite{Altmannshofer:2014pba}, Borexino~\cite{Kamada:2015era,Gninenko:2020xys}, and COHERENT~\cite{AtzoriCorona:2022moj}. Among future proposals, FASER-$\nu$ at CERN aims to search for $\Zpr$ via neutral-current
deep-inelastic neutrino-nucleon scattering~\cite{Cheung:2021tmx}.

Since the $\Zpr$ couples predominantly to second and third generation leptons, the most effective experimental strategy to investigate this model at accelerators is by exploiting muon beams. Dedicated efforts have been proposed at CERN (NA64-$\mu$~\cite{Gninenko:2014pea,Chen:2017awl,Chen:2018vkr,Sieber:2021fue,Kirpichnikov:2021jev}) and Fermilab (FNAL-$\mu$~\cite{Chen:2017awl} and M$^3$~\cite{Kahn:2018cqs}), with NA64-$\mu$ having already completed a pilot run in October 2021. Nevertheless, thanks to the presence of a loop-induced $\Zpr$-electron coupling, $e^-$ beam experiments can also probe a significant portion of the $\Zpr$ parameter space, somehow paving the road to next generation efforts.

In this work, we present the upper limits introduced to the $\Zpr$ parameters space from a search performed with the NA64-$e$ experiment at CERN~\cite{NA64:2019imj}, in the mass range $1-600$ MeV. We consider two $\Zpr$ models, the ``vanilla'' one and a ``dark'' 
scenario in which the $\Zpr$ couples predominantly to light dark sector particles. In the second case, similarly to what was done in Refs.~\cite{Foldenauer:2018zrz,Kahn:2018cqs,Zhang:2020fiu}, we introduce a
dark scalar particle with mass $m_\chi$ and coupling to $\Zpr$ defined by the following Lagrangian:
\begin{equation}
{\cal L}_{D} = g_D \Zpr_\mu J^\mu_D \; \;, 
\end{equation}
where $J^\mu_D$ is the dark vector current given by: 
\begin{align}
    \label{eq:darkcurrent}
    J_{D}^{\mu} = i \left( \chi^* \partial^\mu \chi - \chi \partial^\mu \chi^* \right) \;.
\end{align}
We assume the mass hierarchy $m_\chi < m_\Zpr / 2$ and the couplings ratio $g_D/g_\Zpr \gg 1$. This choice results to a preferred combination of the model parameters that can reproduce the DM relic density observed at present, in the hypothesis that $\chi$ particles are responsible for it:
\begin{equation}\label{eq:bound}
    {g^2_D} g^2_\Zpr \left(\frac{m_\chi}{m_\Zpr}\right)^4 \simeq f \cdot 3 \cdot 10^{-15}\left(\frac{m_\chi}{1\,\mathrm{MeV}} \right)^2\; \; ,
\end{equation}
where $f=1\textendash10$ depends on the nature of the dark sector particle (scalar, Dirac or Majorana fermion, $\ldots$)~\cite{Kahn:2018cqs}. We observe that, since the $L_\mu-L_\tau$ model considered here is associated to a $U(1)$ gauge symmetry, all the $\Zpr$ interactions should be proportional to the gauge coupling $g_\Zpr$. Hence, $g_D$ should be decomposed into a product $g_\Zpr \cdot q_\chi$, where $q_\chi$ is the $\chi$ field charge associated to the new $U(1)$ gauge group.  The ``dark'' scenario considered in this work corresponds to the case $q_\chi \gg 1$, already introduced in Ref.~\cite{Kahn:2018cqs}. For convenience, all the results reported below will be presented as a function of $g_D$.
\\
\\
The paper is organized as follows. In Sec.~\ref{sec:pheno} we introduce the phenomenology for $\Zpr$ production in fixed target electron/positron beam experiments, with a focus on missing-energy efforts. In Sec.~\ref{sec:NA64} we briefly describe the NA64-$e$ experiment at CERN. In Sec.~\ref{sec:methodology} we present our strategy to extend the existing NA64-$e$ results to the $\Zpr$ model, for both the ``vanilla'' case and the ``dark'' one, and finally in Sec.~\ref{sec:results} we show and discuss the obtained results, including the sensitivity projections for future NA64-$e$ runs.

\section{\label{sec:pheno}$\Zpr$ production in fixed target electron-beam experiments}

\begin{figure}[t]
    \centering
    \includegraphics[width=.35\textwidth]{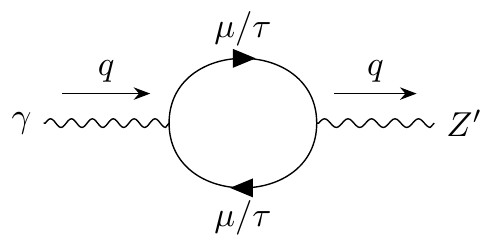}
    \caption{The loop diagram inducing a kinetic mixing between the $\Zpr$ and the photon.}
    \label{fig:Loop}
\end{figure}

\begin{figure}
    \centering
    \includegraphics[width=.45\textwidth]{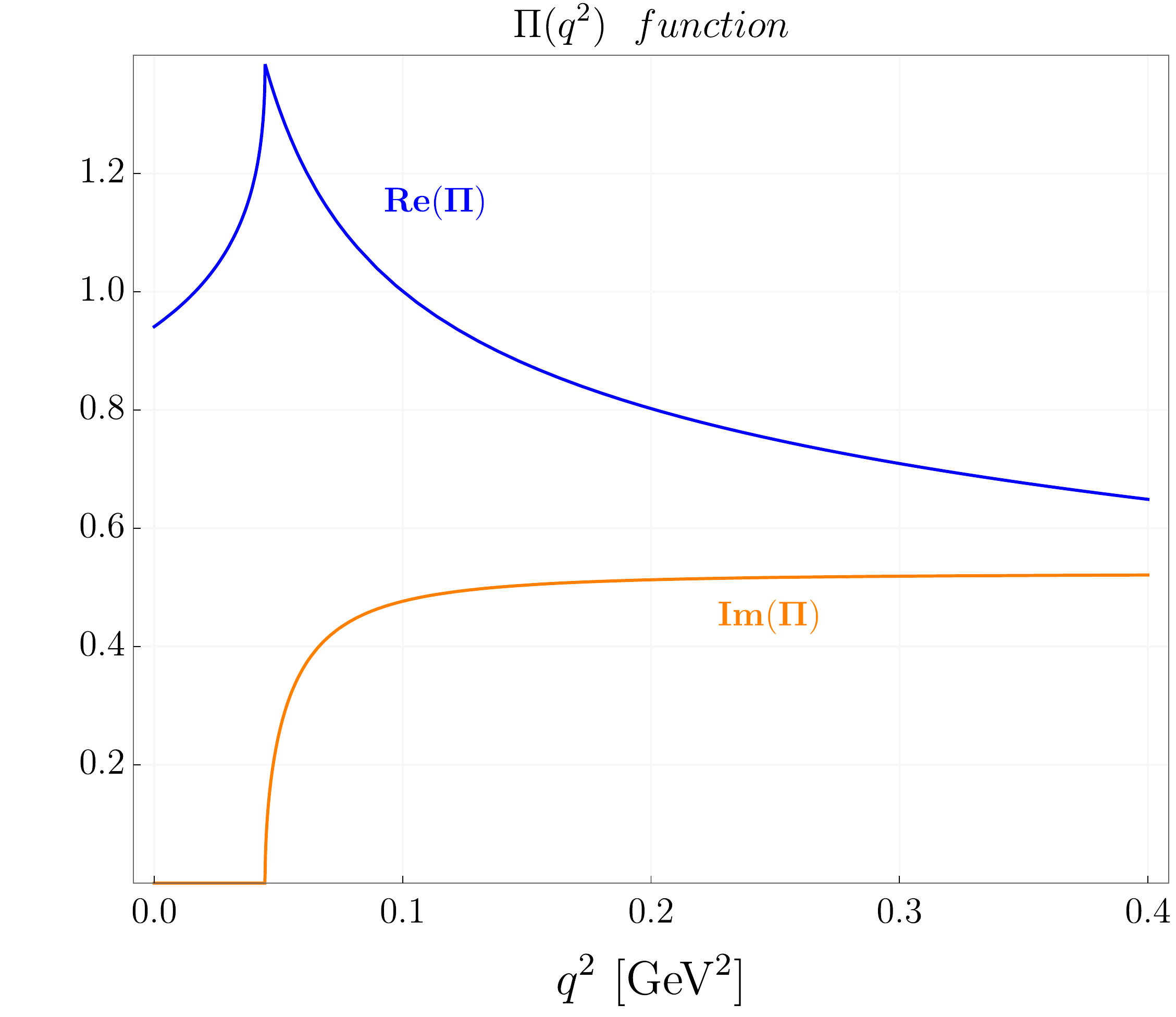}
    \caption{The real and imaginary part of the function $\Pi(q^2)$. For illustration purposes, the arbitrary coupling choice $g_\Zpr=2\pi^2/e$ was made.}
    \label{fig:Pi}
\end{figure}

The $\Zpr$ model considered in this work does not include explicitly a kinetic mixing term between the $\Zpr$ and the SM photon, that would result in a tree-level coupling with the SM electric charge. However, such a coupling arises naturally from the one-loop diagrams reported in Fig.~\ref{fig:Loop}, introducing an effective $e^\pm - \Zpr$ interaction term $e \Pi(q^2) \Zpr_\mu \left(\overline{e} \gamma^\mu e\right)$, where the complex function $\Pi(q^2)$ depends on the momentum $q^2$ carried by the $\Zpr$~\cite{Araki:2017wyg,Gninenko:2018tlp,Zhang:2020fiu}:
\begin{equation}
\label{eq:pi}
    \Pi(q^2)=\frac{e\,g_\Zpr}{2\pi^2}\int_0^1dx \, x (1-x) \ln\frac{m^2_\tau-x(1-x)q^2}{m^2_\mu-x(1-x)q^2}\;\;.
\end{equation}
A plot of the $\Pi(q^2)$ function is reported in Fig.~\ref{fig:Pi}. As already pointed out in Ref.~\cite{Araki:2017wyg}, the dependence of the $e^\pm - \Zpr$ coupling on the momentum is a unique feature of this model, and makes the phenomenology of $\Zpr$ searches at electron/positron beam experiments significantly different than the dark photon case, where the coupling is constant~\cite{Holdom:1985ag}. For very small values of the $\Zpr$ momentum, $q^2\ll m^2_\mu$, the $\Pi(q^2)$ function assumes the constant value $\Pi(0)=\frac{e\,g_\Zpr}{6\pi^2}\ln\frac{m_\tau}{m_\mu}\simeq 0.0144\cdot g_\Zpr$, and the $\Zpr-e^\pm$ interaction resembles that of the ``traditional'' dark photon model under the exchange $\varepsilon \leftrightarrow 0.0144\cdot g_\Zpr$, where $\varepsilon$ is the dark photon kinetic mixing parameter~\cite{Holdom:1985ag,Fabbrichesi:2020wbt}. At larger momentum values, however, there is an enhancement of $\Pi(q^2)$, with a maximum value for $q^2=4m^2_\mu$, where its magnitude is a factor $\sim 1.5$ larger than its small-momentum value, resulting in a significant increase of the $\Zpr$ production yield in this kinematic region. 

\begin{figure}[t]
    \centering
    \includegraphics[width=.45\textwidth]{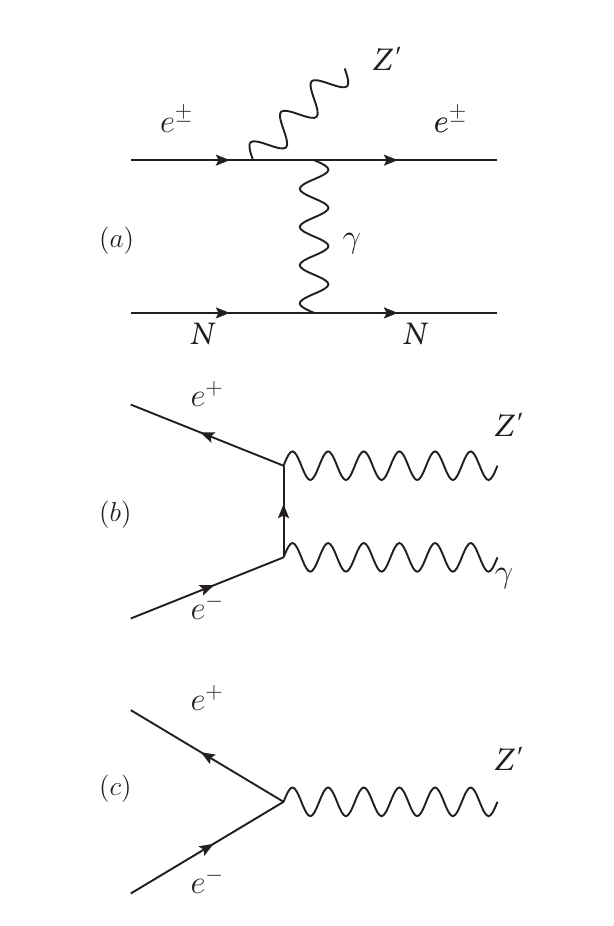}
    \caption{The three main $\Zpr$ production processes for an electron/positron beam impinging on a fixed target: 
  $(a)$~radiative $\Zpr$ production;
  $(b)$~non-resonant $e^+e^-$ annihilation; 
  $(c)$~resonant $\Zpr$ production in $e^+e^-$ annihilation.}
    \label{fig:diagrams}
\end{figure}

The main $\Zpr$ production processes in the collision of a high energy electron or positron beam with a fixed thin target are shown in Fig.~\ref{fig:diagrams}. Diagram (a) corresponds to the so-called $\Zpr$-strahlung process, in which a $\Zpr$ is radiatively emitted by the lepton interacting with the electromagnetic field of a nucleus in the target. Diagrams (b) and (c), relevant only for an impinging positron, correspond to the non-resonant (b) and resonant (c) $e^+e^-$ annihilation. These processes are analogous to those relevant for the production of a dark photon, with the significant difference of the loop-induced $\Pi(q^2)$ factor appearing in the effective coupling. The corresponding production cross-section $\sigma_P(E,m_\Zpr)$ formulas, considering an on-shell $\Zpr$ and a beam energy $E$, can be obtained from the corresponding expressions for a dark photon model (see, e.g., ~\cite{Bjorken:2009mm,Marsicano:2018glj,Marsicano:2018krp}) with the substitution $\varepsilon \leftrightarrow \Pi(m^2_\Zpr)$. Depending on the model and on the specific parameter values, the produced $\Zpr$ can decay to different final states. For the ``dark'' case, in the mass range $2m_\mu < m_\Zpr < 2m_\tau$, the following decay channels are possible (neglecting the strongly suppressed $\Zpr \rightarrow e^+e^-$ decay):
\begin{align}
    \Gamma(\Zpr \rightarrow \nu \overline{\nu})=&
    \frac{\alpha_{\Zpr}}{3}m_\Zpr\\ 
    \Gamma(\Zpr \rightarrow \mu^+ \mu^-)=&
    \frac{\alpha_{\Zpr}}{3}m_\Zpr\left(1+2r^2_\mu\right)\sqrt{1-4r^2_\mu}\\
    \Gamma(\Zpr \rightarrow \chi\chi)=&\frac{\alpha_D}{12}m_\Zpr\left(1-4r^2_\chi\right)^{\frac{3}{2}} \; \; ,
\end{align}
where $\alpha_\Zpr\equiv \frac{g^2_\Zpr}{4\pi}$, $\alpha_D\equiv \frac{g^2_D}{4\pi}$, $r_\mu\equiv m_\mu/m_\Zpr$, $r_\chi\equiv m_\chi/m_\Zpr$, and the neutrino channel refers to the summed contributions from $\nu_\mu$ and $\nu_\tau$. The ``vanilla'' scenario results can be simply obtained by setting $g_D=0$. 
We observe that, for $m_\Zpr < 2m_\mu$, only the invisible decay channels to neutrinos, and eventually to dark sector particles, are allowed.

For a thin target electron-beam experiment, with $t\ll X_0$, where $t$ is the target thickness and $X_0$ the radiation length, only diagram $(a)$ contributes to the $\Zpr$ event yield, scaling as:
\begin{equation}
   N_S \propto t \int d\vec{x} \frac{d\sigma_{Rad}(\vec{x})}{d\vec{x}} \; \; ,
\end{equation}
where $\vec{x}$ denotes a set of kinematic variables to describe the final state phase space.

In the case of a thick target electron-beam experiment with $t \gg X_0$, all the aforementioned production channels contribute to the $\Zpr$ yield due to the presence of the secondary electrons and positrons in the electromagnetic shower
induced by the primary electron.
 For the specific case of a missing-energy experiment, in which the signal is associated to the production of invisibly-decaying $\Zpr$ particles with energy greater than a threshold $E_{cut}^{Miss}$, the event yield scales as~\cite{Izaguirre:2014bca,Marsicano:2018glj}:
\begin{align}
N_{S} \propto & 
\int dE \, dE_F \,T_\pm(E)\frac{d{\sigma}_{Rad}(E,E_F)}{dE_F} \;+\\ \nonumber
+&Z\cdot\int dE\, T_+(E)\sigma_{Res}(E)\;\;,
\end{align}
where $T_-$ ($T_+$) is the secondary electrons (positrons) differential track-length distribution~\cite{Chilton,Tsai:1966js} as a function of their energy $E$, and $T_\pm \equiv T_- + T_+$. 
The quantity $\frac{d\sigma_{Rad}(E_\pm,E_F)}{dE_F}$ is the differential cross section \textit{per nucleus} for radiative $\Zpr$ production with respect to the final state invisible particles total energy $E_F$, while $\sigma_{Res}$ is the total cross section \textit{per electron} for the resonant production process. $Z$ is the atomic number of the target material\footnote{The cross section $\frac{d\sigma_{Rad}(E_\pm,E_F)}{dE_F}$ contains an implicit quadratic dependence on the atomic number.}. The two integrals in the radiative contribution term are performed over the $E_F > E_{cut}^{Miss}$ range. For the resonant production the kinematic constraint $E_F=E$ reduces the dimensions of the integral region. We did not include the non-resonant $\Zpr$ production mechanism in this computation, since for a primary electron beam the contribution to the total yield due to the 
annihilation with secondary positrons (see e.g. Fig.~\ref{fig:diagrams}(b)) is negligible~\cite{Marsicano:2018glj}.

\section{\label{sec:NA64} The NA64-$e$ experiment}
\begin{figure*}[t]
    \centering
    \includegraphics[width=.8\textwidth]{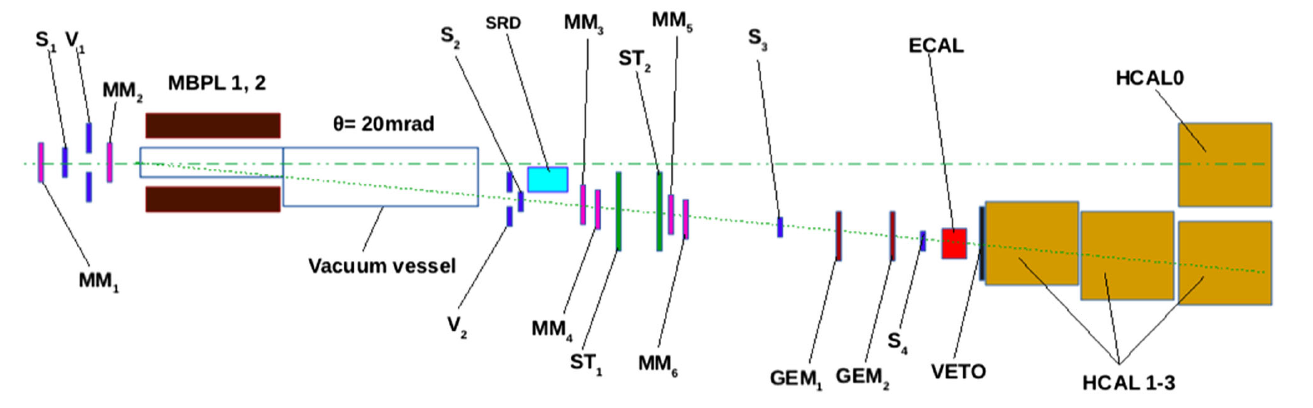}
    \caption{Schematic view of the NA64-$e$ detector in the nominal, invisible mode configuration. See text for further details.}
    \label{fig:detector}
\end{figure*}
The NA64-$e$ experiment at CERN is devoted to the search for dark sector particles feebly interacting with electrons. NA64-$e$ exploits the 100 GeV high-purity, low-current electron beam from the H4 beamline to perform the search, by measuring event-by-event the energy deposited in a thick active target, looking for events with large missing energy (see  Refs.~\cite{Gninenko:2016kpg,NA64:2019imj,Gninenko:2013rka,Banerjee:2017hhz} for a complete description of the detector and of the missing-energy approach). A schematic view of the detector is shown in Fig.~\ref{fig:detector}. The NA64-$e$ active thick target is an inhomogeneous electromagnetic calorimeter (ECAL), with energy resolution $\sigma_{E}/E\simeq 10\%/\sqrt{E\mathrm{(GeV)}}+4\%$. The ECAL is made by 150 alternated layers of 1.5-mm thick lead plates and 1.5-mm thick plastic scintillator tiles, for a total length of about 40 $X_0$; the detector is segmented into a 6x6 matrix of independent transverse cells, with each cell further divided into a 4 $X_0$ pre-shower section and a main section.

Two main types of backgrounds, resulting to missing energy events, affect the NA64-$e$ experiment. The first type is associated to the production of one or more penetrating particles in the ECAL by the primary beam. To suppress this contribution, a massive hadronic calorimeter (HCAL) is installed downstream of the ECAL. A high-efficiency plastic scintillator detector (VETO) is also located between the two calorimeters, to identify events in which penetrating charged particles are produced. The second type of background events is due to residual $\approx 1\%$ hadron contaminants in the primary beam. To suppress these, a syncrotron radiation beam-tagging system (SRD) is installed upstream of the ECAL~\cite{Depero:2017mrr}. The NA64-$e$ detector assembly also includes a magnetic spectrometer to measure the momentum of impinging particles. This consists of two successive dipole magnets (total magnetic strength $\int B dl \simeq 7$ T$\cdot$ m) and a set of upstream and downstream tracking detectors, Micromegas (MM), Strawtubes (ST) and Gaseous Electron Multipliers (GEM). Finally, a set of  beam-defining plastic-scintillator counters (SC) is present.
During operations, the majority of the primary electrons gives rise to an electromagnetic shower in the ECAL, with full energy release in the detector. To suppress the rate of events processed and written to disk, the experiment trigger requires, other than a coincidence signal between the SCs, that the total energy sum signal from the ECAL corresponds to an energy deposition of less than 80 GeV.

NA64-$e$ already completed data-taking campaigns in 2016, 2017, 2018 with a total accumulated charge of about $N_{EOT}=2.84 \cdot10^{11}$ electrons-on-target (EOT). The selection criteria adopted in the analysis were identified by optimizing the experiment sensitivity, adopting a blind-analysis approach~\cite{Banerjee:2017hhz,NA64:2019imj,Andreev:2021fzd}. These include the requirement to have a well reconstructed track in the upstream spectrometer, with momentum in the range~$(100\pm3)$ GeV, an in-time cluster in the SRD detector, and a shower signal in the ECAL with the longitudinal and transverse shape of a missing-energy event. The latter selection also included a 0.5 GeV energy cut for the ECAL pre-shower section. After applying all selection cuts, no events were observed in the signal region, defined by the two requirements $E_{ECAL}<50$ GeV and $E_{HCAL}<1$ GeV; this observation is compatible with the estimate of $(0.53\pm0.17)$ background events, mostly due to the interaction of electrons with upstream beamline elements, producing a soft electron hitting the ECAL and one or more hadrons at large angle missing the NA64-$e$ detector. This result was used to set an exclusion limit for the production of an invisibly decaying dark photon ($\Apr$), taking into account both the  radiative and resonant production~\cite{Andreev:2021fzd}. 

\section{\label{sec:methodology} Methodology}
The analysis presented in this work is based on the dataset already scrutinized by the NA64 collaboration to set limits for an invisible decaying dark photon, preventing us to adopt a blind approach. For this work, we decided to follow a strategy based on the same selection criteria adopted in the aforementioned analysis, including the signal region definition. Our approach is based on the observation that, for an electron beam missing energy experiment,  the $\Zpr$ signal would differ from the $\Apr$ one only due to the momentum dependence of the $e-e-\Zpr$ vertex, associated to the $\Pi(q^2)$ function.

We considered first the simpler case in which only the radiative $\Zpr$ production channel is included. The 90$\%$ C.L. upper limit $\varepsilon^{up}$ for the dark photon kinetic mixing parameter reported by NA64-$e$ in Ref.~\cite{NA64:2019imj}, given the negligible number of expected background events, corresponds to an expected number of signal events equal to $N_{up}\simeq2.3$ via the relation:
\begin{equation}
    N_{up} = (\varepsilon^{up})^2 \, \mathcal{N} \, \int dE dE_F \, T_\pm(E) 
    \frac{d\sigma_{\Apr,Rad}}{dE_F} \, \eta_\Apr \; ,
\end{equation}
where $\frac{d\sigma_{\Apr,Rad}}{dE_F}$ is the differential $\Apr$ production cross section \textit{per nucleus} divided by $\varepsilon^2$, $\eta_\Apr$ is the corresponding signal acceptance and detection efficiency, and $\mathcal{N}$ is the overall normalization factor, accounting for the total accumulated number of EOT and for the detector material composition. Due to the detector geometry, $T(E)$ is almost the same for the plastic scintillator and the lead, while a $Z^2$ dependence is included in the cross section: since $\mathcal{N}$ scales as the material density over the atomic mass, in the following we will consider only the contribution from the lead (the same approximation was made in the analysis presented in Ref.~\cite{NA64:2019imj,Andreev:2021fzd}). A similar relation holds for the $\Zpr$ case:
\begin{equation}\label{eq:R}
    N_{up} = (g^{up}_\Zpr)^2 \, \mathcal{N} \, \int dE dE_F \, T_\pm(E) 
   \frac{d{\sigma}_{\Zpr,Rad}}{dE_F}\, \eta_\Zpr \; \;.
\end{equation}
Here we did not show explicitly the integration over the invariant mass $q^2$ of the $\Zpr$ decay particles in the final state, where the $\Pi(q^2)$ dependence enters (see also the Appendix).
By taking the ratio of these expressions, the following expression is obtained:
\begin{equation}\label{eq:limit}
     (g^{up}_\Zpr)^2 = (\varepsilon^{up})^2\, \frac{
    \int dE dE_F\, T_\pm(E) \frac{d\sigma_{\Apr,Rad}}{dE_F} \, \eta_\Apr 
  }
  { \int dE dE_F \, T_\pm(E)\frac{d\sigma_{\Zpr,Rad}}{dE_F} \,\eta_\Zpr
   }
   \equiv (\varepsilon^{up})^2 \mathcal{R} \; \;.
\end{equation}
In this ratio, any absolute normalization factor appearing in the predicted signal yield, such as the EOT number, cancels out, drastically simplifying the calculation. The same is true for all contributions to the signal efficiency terms $\eta_\Apr$ and $\eta_\Zpr$ that are almost independent from the signal model, as discussed in Sec.~\ref{sec:acceptance}. We also observe that this procedure does not depend on the specific value of $N_{up}$, that can be actually slightly different from the ``nominal'' setting (2.3 events) due to the inclusion of the non-zero number of expected background events and of the systematic uncertainty factors in the statistical procedure.

To also include the $\Zpr$ resonant production channel in the upper limit calculation, we modified the denominator appearing in the definition of $\mathcal{R}$ adding the $e^+e^- \rightarrow \Zpr \rightarrow invisible$ event yield
to the total:
\begin{equation}\label{eq:limit2}
\begin{split}
     \int dE dE_F & \, T_\pm(E)\frac{d{\sigma}_{\Zpr,Rad}}{dE_F} \,\eta_\Zpr \longrightarrow \\
     \longrightarrow &
     \int dE dE_F \, T_\pm(E)\frac{d{\sigma}_{\Zpr,Rad}}{dE_F} \,\eta_{\Zpr,Rad} +\\
     +
     &Z\int dE\, T_+(E){\sigma}_{\Zpr,Res}\,\eta_{\Zpr,Res} \; \;.
\end{split}
\end{equation}

To solve Eq.~\ref{eq:limit}, we observe that $\mathcal{R}$ still contains a residual dependence on $g_\Zpr^{up}$ due to the total $\Zpr$ width $\Gamma_\Zpr$ and the additional factor $g_\Zpr^2$ for the $e^+e^-\rightarrow \Zpr \rightarrow \nu\overline{\nu}$ channel (see also the Appendix). To account for this we proceeded by iteration, starting from the \textit{ansatz} $g^{up}_\Zpr = \left(g^{up}_\Zpr\right)_0$, where $\left(g^{up}_\Zpr\right)_0$ was obtained from the narrow-width approximation (``vanilla'' case) or considering the decay to dark sector particles only (``dark'' case). At each $n-$th iteration, we used the value $\left(g^{up}_\Zpr\right)_{n-1}$ to compute $\Gamma_\Zpr$ and obtain  $\left(g^{up}_\Zpr\right)_{n}$ via Eq.~\ref{eq:limit}. Convergence was observed already after two iterations.

\subsection{Electrons and positrons track-length}

We computed the electrons and positrons track-length in the NA64-$e$ ECAL through a Monte Carlo simulation, exploiting the NA64-$e$ GEANT4-based framework~\cite{Agostinelli:2002hh}. The full NA64-$e$ detector geometry and material composition were implemented in the simulation, including the magnetic field bending the impinging 100 GeV electron beam. Primary electrons were generated just before the upstream tracking stations, with a beam spot size of 1.5 cm and an angular divergence of 0.1 mrad.
For all electrons and positrons propagating in the ECAL volume, we sampled the particle energy at each discrete step the trajectory is divided into by GEANT4. We then constructed the corresponding $e^-$ and $e^+$ energy distributions in the lead and in the plastic scintillator, by assigning to each sampled value a weight given by the step length. The electrons (positrons) track length $T_-(E)$ ($T_+(E)$) was finally obtained by normalizing by the total number of simulated events.
\begin{figure}[!t]
    \centering
    \includegraphics[width=.48\textwidth]{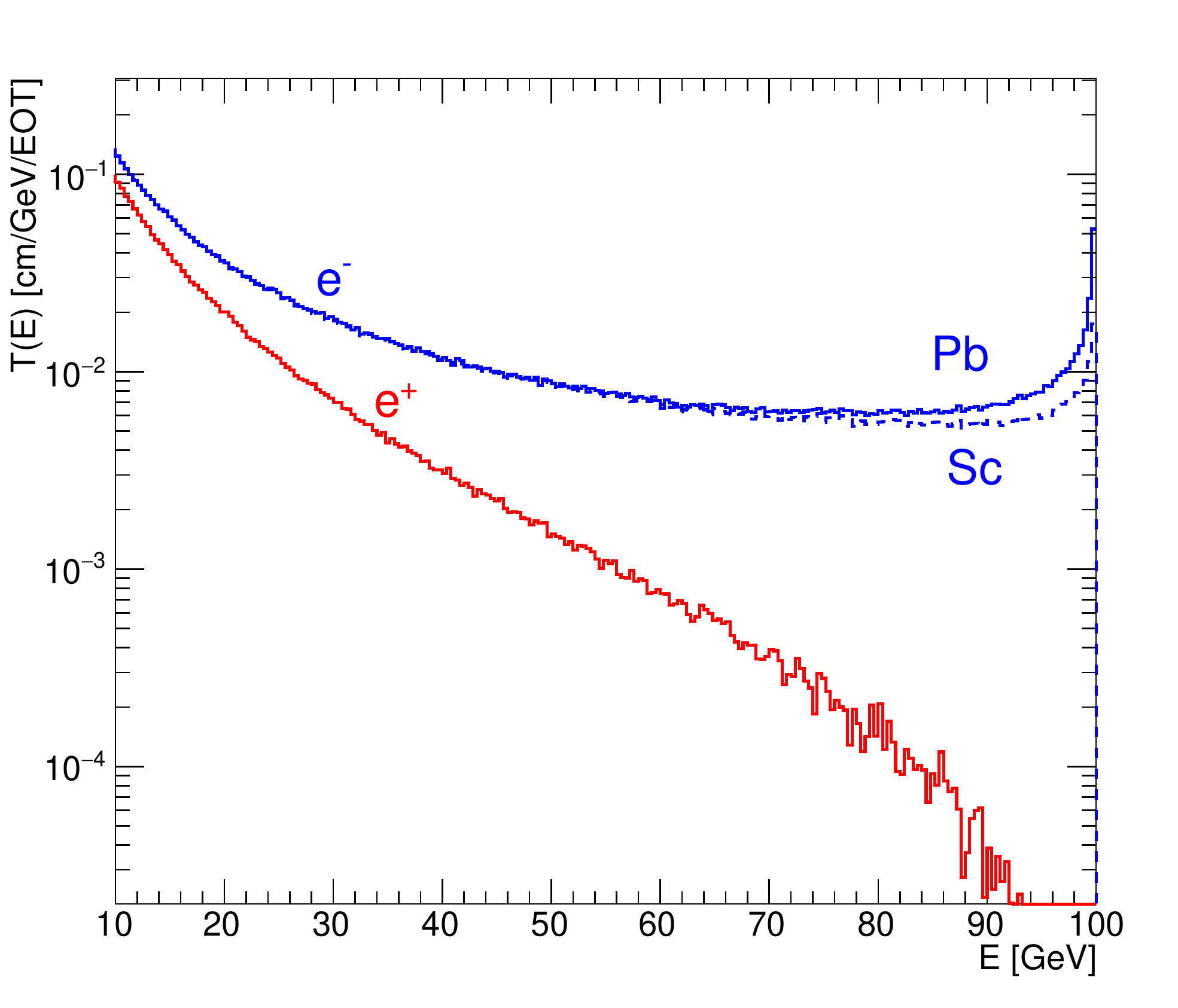}
    \caption{The electrons and positrons differential track length in the NA64-$e$ ECAL as a function of the energy. For positrons, the two distributions for lead and plastic scintillator are almost identical.}
    \label{fig:TL}
\end{figure}
The obtained result is reported in Fig.~\ref{fig:TL}, displaying the electrons and positrons track length in the lead and in the plastic scintillator. We observe that, due to the ECAL segmentation into equally-sized layers of these materials, the track lengths are almost identical. The factor $\simeq 2$ difference for the high-energy part of the $e^-$ distribution is due to the fact that in each layer, including the first, the lead is located in front of the scintillator. Therefore, the energy of the electrons propagating into the first scintillator tile is systematically smaller than that in the first lead layer.

Since, by default, GEANT4 forces a new trajectory step every time a particle crosses the boundary between two regions, we exploited the intrinsic 1.5-mm longitudinal segmentation of the NA64-$e$ ECAL cells to ensure a proper track-length evaluation, without imposing any further subdivision of the particles trajectory. The consistency of the result regarding this choice was checked by repeating the computation of $T(E)$ enforcing a maximum step length of 0.50-mm in the simulation. 
We observed no significant variations for $T_+(E)$. For the $T_-(E)$, instead, the two distributions are almost equivalent up to $E \simeq 80$ GeV, while a difference of up to $20\%$ is observed for $80<E<99.5$ GeV with larger values predicted by the 0.5-mm maximum step-length simulation. For $E>99.5$ GeV the difference is even higher, reaching a factor up to 5. The overall normalization of the two distributions is equivalent. 
We explained this as being related to the aforementioned ECAL geometry. A single 1.5-mm primary electron step in the first lead layer would contribute to the $T_-(E)$ sampling with an intermediate $E$ value, whereas if the same step was divided into three 0.5-mm segments, the first would increment the high-energy portion of $T_-(E)$. The same effect also applies, in general, for the first ECAL thicknesses, with a reduced intensity. In conclusion, considering that the track-length normalization is not affected by the choice of the stepping size, and that the radiative $\Zpr$ emission has a smooth dependence on the beam energy, in the following we will use the $T_\pm(E)$ result obtained from the nominal GEANT4 simulation.

\subsection{Signal acceptance and detection efficiency}\label{sec:acceptance}

The NA64-$e$ signal efficiency for the $\Apr$-strahlung channel $\eta_\Apr$, the $\Zpr$-strahlung channel $\eta_{\Zpr,Rad}$ and the $\Zpr$ resonant production $\eta_{\Zpr,Res}$ can be factorized into two different terms. 
The first, $\eta^{up}$, is associated to the response of all detector components installed \textit{upstream} the NA64-$e$ ECAL. This term thus include the tracking efficiency, the efficiency of the SRD cut to reject beam contaminants, as well as the efficiency of the SC counters included in the trigger condition. Global effects such as the overall DAQ efficiency can also be included in this term. The second term $\eta^{down}$, instead, is associated to the ECAL, the VETO, and the HCAL detector responses - the main contribution being the 50 GeV ECAL missing energy threshold. By definition, $\eta^{up}$ is the same for all reaction channels and thus cancels out in the definition of $\mathcal{R}$. Since these channels are characterized by a different signal kinematics, instead, $\eta^{down}$ has to be computed specifically for each of them.

We evaluated $\eta^{down}$ for the different reaction channels and parameter models through a GEANT4 simulation of the full NA64-$e$ detector setup using the DMG4 package for events generation~\cite{Bondi:2021nfp} and adopting an ad-hoc cross-section biasing mechanism to enhance signal production without distorting the corresponding kinematics, similarly to what was performed in Ref.~\cite{Andreev:2021fzd}.
The DMG4 package does not offer the possibility to consider off-shell $Z'$ production, therefore we used the on-shell approximation $\Pi(q^2) \rightarrow \Pi(m^2_\Zpr)$ in the calculation of $\eta^{down}$; this is justified by the fact that the NA64 detector is sensitive to the missing energy and not to the kinematics of the invisible decay particles. The Monte Carlo event samples were processed through the same NA64-$e$ reconstruction code used for the data analysis, and $\eta^{down}$ was determined from the fraction of these satisfying all the selection cuts associated to the ECAL, the VETO, and the HCAL, possibly as a function of one or more kinematic observables. 

\begin{figure}[t]
    \centering
    \includegraphics[width=.48\textwidth]{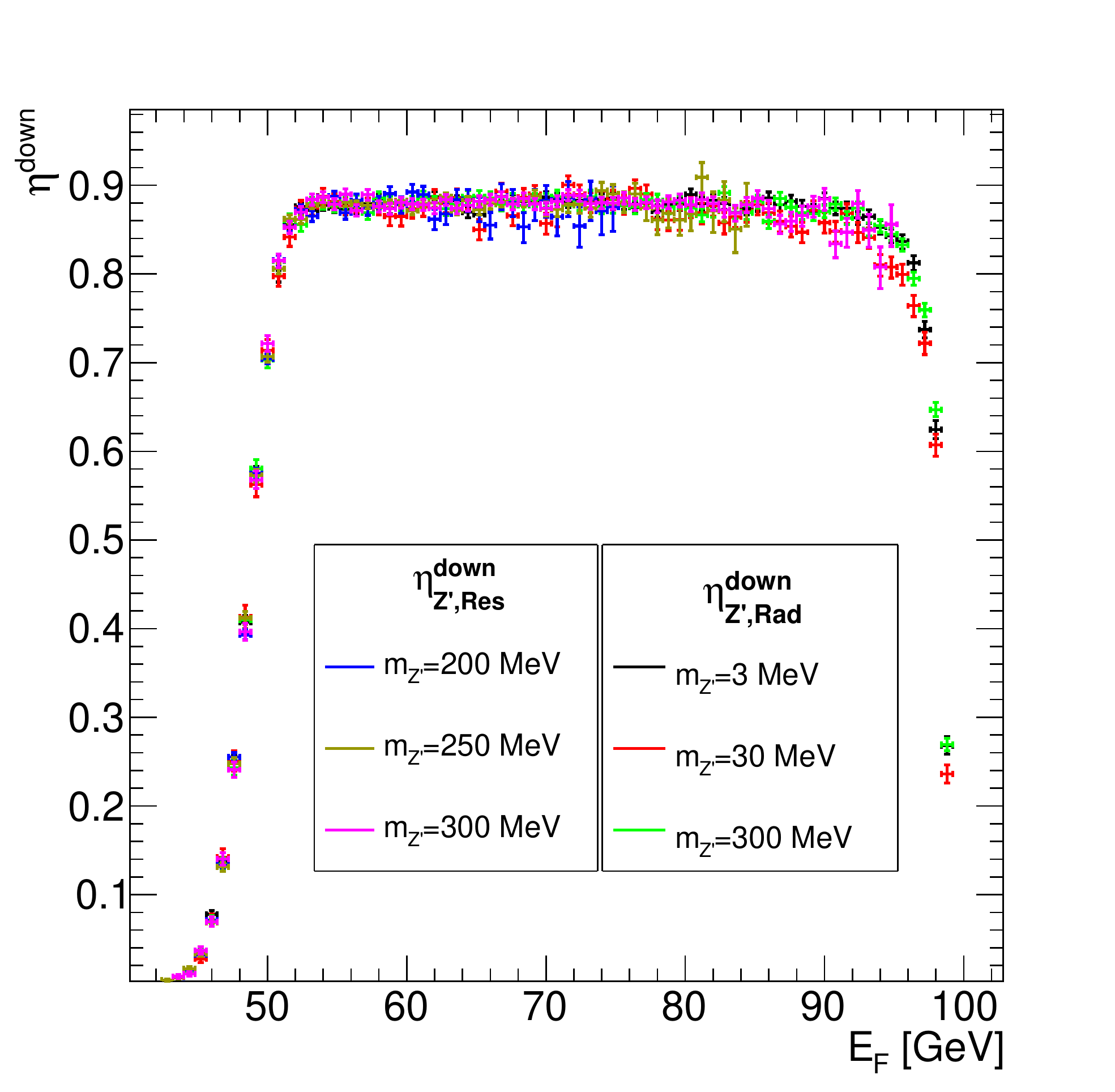}
    \caption{The ``downstream'' NA64-$e$ signal detection efficiency $\eta^{down}$ as a function of the emitted $\Zpr$ energy $E_F$, for different models and corresponding parameters.}
    \label{fig:efficiency}
\end{figure}

The signal efficiency $\eta^{down}$ as a function of the $\Zpr$ energy $E_F$ is shown in Fig.~\ref{fig:efficiency}, for the resonant process at $m_\Zpr=$ 200 MeV, 250  MeV, and 300 MeV, and for the radiative process at $m_\Zpr=$ 3 MeV, 30  MeV, and 300 MeV -- these values are representative of the $\Zpr$ mass range explored in this work. We observe that, at fixed $E_F$, all results are compatible with each other within the errors, thus suggesting that kinematic dependence of $\eta^{down}$ can be effectively taken into account considering its shape as a function of $E_F$. We also checked that the $\eta_{\Apr}^{down}\simeq \eta_{\Zpr,Rad}^{down}$. Therefore, in the following we will use a common expression of $\eta^{down}(E_F)$ for all reaction channels. The smooth transition observed around $E_F=50$ GeV is due to the convolution between the 50 GeV threshold on the energy deposited in the ECAL and its finite resolution. Therefore, by including the energy dependence of $\eta^{down}$ in $\mathcal{R}$, we effectively take into account the modification to the $\Zpr$ line shape due to the detector effects, particularly important in case of resonant production with resonant energy close to the threshold value, i.e. for $m_\Zpr \simeq \sqrt{2m_e E_{cut}^{Miss}} \simeq 225$~MeV. 

\subsection{$\Zpr$ events yield}

We used the \texttt{MADDUMP} event generator to simulate the $\Zpr$ production in the NA64-$e$ ECAL~\cite{Buonocore:2018xjk}. \texttt{MADDUMP} is a plugin for the \texttt{MadGraph5\_aMC}$@$\texttt{NLO} program~\cite{Alwall:2011uj,Alwall:2014bza} developed for fixed thick-target setups that allows to compute the differential yield of $\Zpr$ particles in the lead material of the NA64-$e$ ECAL from the knowledge of electrons and positrons differential track length. For the radiative emission process, we adopted the nuclear form-factor parameterization reported in Ref.~\cite{Bjorken:2009mm}. We also explicitly included the factor $\Pi(q^2)$ in the $e-e-\Zpr$ vertex, setting $g_\Zpr=1$ as justified before. For simplicity, we used an effective polynomial interpolation of the full calculation result presented in Sec.~\ref{sec:pheno} -- to account for the cusp at $q^2=4m^2_\mu$, this was implemented separately for the low and high momentum region. Further details are provided in the Appendix.

For a given reaction channel, \texttt{MADDUMP} provides both an unweighted set of $N_{MC}$ Monte Carlo events and the value of the energy-dependent total cross section integrated over the track-length distribution. To include the downstream signal acceptance and detection efficiency, and to account for the ECAL resolution, for each event we computed $\varepsilon_{down}(E_\Zpr)$, summed all these values, and normalized the sum to $N_{MC}$, finally multiplying the integrated cross section by the result. We repeated the calculation independently for the $\Zpr$-strahlung on the lead nucleus target and for the $\Zpr$ resonant annihilation on atomic electrons. By fixing $\Pi(q^2)=1$, we also simulated the radiative dark photon emission on the lead material, necessary to compute the $\mathcal{R}$ numerator in Eq.~\ref{eq:limit}.

\section{\label{sec:results} Results}

\begin{figure*}[t]
    \centering
    \includegraphics[width=.48\textwidth]{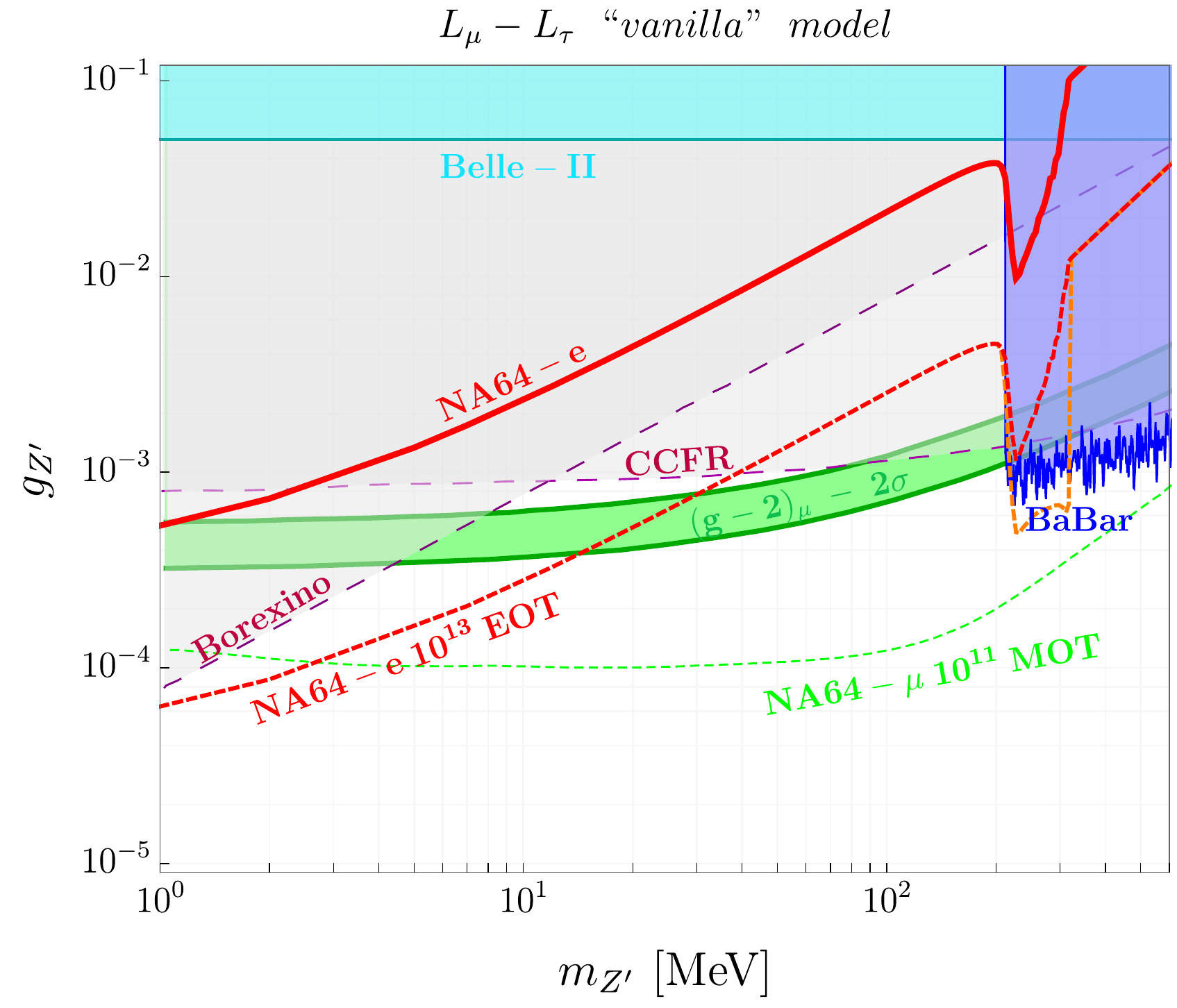}
    \includegraphics[width=.48\textwidth]{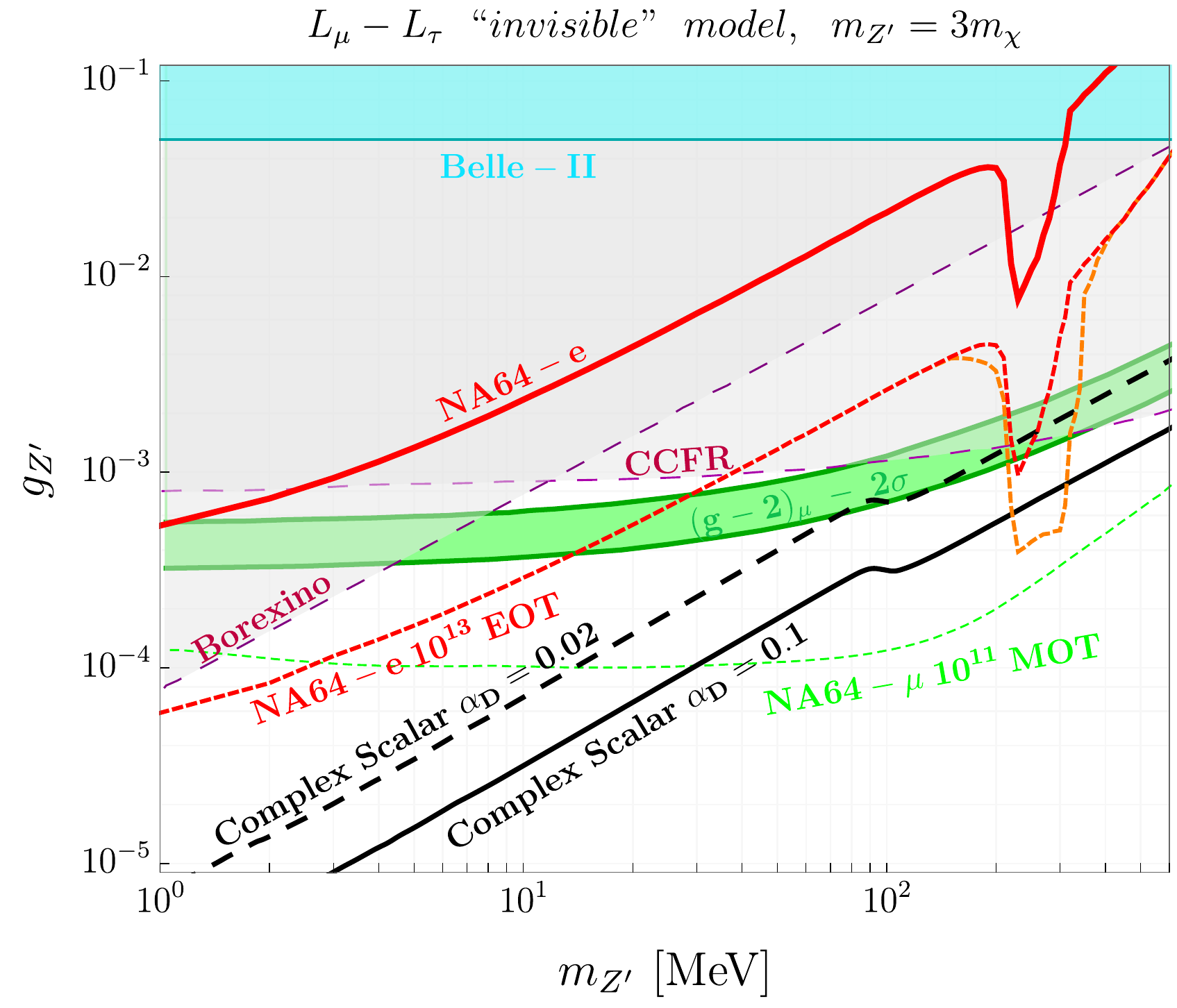}
    \caption{
    The NA64-$e$ exclusion limit for the $L_\mu-L_\tau$ model, for the ``vanilla'' (left) and ``dark'' (right) flavour (red curve). 
   The red (orange) dashed curves represent the sensitivity projections for a future high-statistics NA64-$e$ run with an electron (positron) beam, for a total accumulated charge of $10^{13}$ EOT, while the green dashed curve is the sensitivity projection of NA64-$\mu$~\cite{Sieber:2021fue}.
    The gray areas are the regions excluded by phenomenological re-analysis of neutrino experiments~\cite{Altmannshofer:2014pba,Kamada:2015era}, while the blue region is the area excluded by BaBar~\cite{BaBar:2016sci} for the ``vanilla'' case.
    Finally, the black curves represent the so-called ``thermal target'' for the two values of $\alpha_D=0.1$ and $\alpha_D=0.02$, i.e. the preferred combination of the parameters to explain the observed dark matter relic density. These have been calculated through Eq.~\ref{eq:bound} by re-scaling the results from Ref.~\cite{Kahn:2018cqs}. }
    \label{fig:result1}
\end{figure*}

The 90$\%$ CL exclusion limits in the $m_\Zpr$ vs $g_\Zpr$ parameter space obtained from the NA64-$e$ experiment are shown in Fig.~\ref{fig:result1}, for the ``vanilla'' model (left panel) and for the ``dark'' one (right panel). In the latter case, to check the effect of changing the $\Zpr$ width, we considered  the dark coupling values $\alpha_D=0.1$ ($g_D=1.1$) and $\alpha_D=0.02$ ($g_D=0.5$), with the fixed mass ratio $m_\Zpr/m_\chi=3$. Since, for these values of $\alpha_D$, the missing energy resolution of the ECAL is larger than the $\Zpr$ width, no significant differences are observed between the two cases. Due to tension with perturbative unitarity bound, larger $\alpha_D$ values were not considered~\cite{Kahn:2018cqs}. The shape of the upper limit curve is associated to the diagram that mostly contributes to the signal yield for a given  $m_\Zpr$ value. In the region where it is kinematically allowed, roughly defined by $E_{cut}^{Miss}<m^2_\Zpr/(2m_e)<E_0$, the resonant emission dominates, resulting in the peculiar ``cusp'' visible at $m_\Zpr\simeq 250$ MeV, while for other $m_\Zpr$ values the signal yield is entirely due to the radiative process. The differences between the limit curves in the ``vanilla'' one and ``dark'' scenarios are concentrated in the region were the resonant production dominates ($\sim 230$ MeV $-330$ MeV). In this mass range the convolution of the $\Pi(q^2)$ function over the $\Zpr$ line shape results in a slight widening of the resonant ``cusp'' in the ``dark'' scenario with respect to the ``vanilla'' one.

We underline that our procedure guarantees that the obtained limit accounts for the effect of all systematic uncertainties that were included in the NA64-$e$ $\Apr$ search analysis (see Ref.~\cite{NA64:2019imj}). We evaluated the effect of using a global function for $\eta^{down}(E_F)$ by repeating the calculation of $\mathcal{R}$ using separately the expressions for $\eta^{down}_{\Zpr,Res}$, $\eta^{down}_{\Zpr,Rad}$, and $\eta^{down}_{\Apr,Rad}$ for the mass value $m_\Zpr=250$ MeV, where the resonant contribution is dominant. The obtained result is compatible within $1\%$ with the previous one.

 We report in the same figure the constraints from other accelerator-based experiments, namely the BaBar search through the visible decay $\Zpr \rightarrow \mu^+\mu^-$~\cite{BaBar:2016sci} and the Belle-II invisible search result~\cite{Belle-II:2019qfb}. For the ``dark'' scenario the BaBar limit, obtained from a search exploiting the $\Zpr\rightarrow \mu^+\mu^-$ decay, does not apply, since for $g_D\gg g_\Zpr$ the $Z^\prime$ decays mostly to the invisible $\chi\chi$ channel. In the same plots, together with the preferred ``band'' from the muon $g-2$ anomaly, we also show results obtained by different authors through a re-analysis of data reported by neutrino experiments, namely the CCFR result for the trident $\nu N \rightarrow \nu N \mu^+\mu^-$ production and the Borexino measurement of solar $^{7}$Be $\nu_e$ scattering on atomic electrons -- however, we point out that these results should be somehow considered \textit{cum grano salis}, since in both cases not all the experimental details of the original measurement where taken into account in the re-analysis, for example the detector energy resolution; also, the theoretical assumptions for the Borexino limit were questioned in Ref.~\cite{Gninenko:2020xys}, and a 30$\%$ discrepancy was found. The bound obtained from the re-analysis of the COHERENT data, not reported in the figure, is less stringent than those from CCFR and Borexino~\cite{AtzoriCorona:2022moj}. We included the sensitivity projection for NA64-$\mu$, a parallel effort of the NA64 collaboration. NA64-$\mu$ is a missing-momentum experiment at the CERN M2 beamline, employing the 160 GeV muon beam from SPS to search for the $\Zpr$ via the reaction $\mu N \rightarrow \mu N \Zpr$ and the subsequent invisible $\Zpr$ decay. As discussed in Ref.~\cite{Sieber:2021fue}, the sensitivity curve has been computed for an accumulated statistics of 10$^{11}$ muons-on-target (MOT), for which zero background events are expected. We also report the sensitivity projection for NA64-$e$ for a future high statistics run of $10^{13}$ EOT, assuming the same run conditions of the current $e^-$-beam dataset and considering zero background events. The NA64 collaboration is also  investigating the possibility to perform a missing energy experiment with a positron beam, to maximize the signal yield induced by the $e^+e^-$ channel: we thus show the sensitivity projection for a $10^{13}$ positrons-on-target experiment, again considering zero background events. This result has been obtained following the same procedure used for the electron beam analysis. The track-length and the efficiency were evaluated via \texttt{GEANT4} and the cross section was numerically integrated with \texttt{MADDUMP}.

The continuous and dashed black curves represent the ``thermal target'', i.e. the preferred combination of the parameters to explain the observed dark matter relic density, calculated through Eq.~\ref{eq:bound} by re-scaling the results from Ref.~\cite{Kahn:2018cqs}.



\section{\label{sec:conclusion} Conclusions}

The $L_\mu-L_\tau$ gauge symmetry extension of the Standard Model provides an elegant explanation to observed ``anomalies'' between data and SM predictions, such as the muon magnetic moment puzzle. In the ``dark'' flavour, the corresponding $\Zpr$ gauge boson acts as a portal between SM and a ``dark sector'', possibly connected with the DM phenomenology. In this work, we presented the exclusion limits for the $\Zpr$ parameters space obtained from the analysis of the existing NA64-$e$ experiment dataset, based on a rigorous treatment of the momentum-dependent coupling between the $\Zpr$ and the first-generation leptons induced by a loop mechanism. These are the first limits set by a direct experimental search for $\Zpr$ that exclude the region up to the muon $g-2$ preferred band for $m_\Zpr \simeq 1$ MeV,  confirming the results already reported by the re-interpretation of neutrino experiments data. Our work demonstrates the potential of the NA64$-e$ experiment, also regarding the complementarity to the future searches with NA64$-\mu$. Future high-statistics NA64-$e$ runs will explore even larger regions in the parameters space, with the positron-beam measurement playing a significant role due to the electron-positron annihilation production mechanism.

\begin{acknowledgments}
AC and LM warmly thanks Luca Buonocore for very helpful support regarding \texttt{MADDUMP}. AC also thanks Patrick Foldenauer for the useful discussion concerning the connection between the $L_\mu-L_\tau$ model and the dark matter phenomenology and Bertrand Echenard for providing the numerical values for the existing BaBar exclusion  limits. We warmly thank the reviewer of the manuscript for providing us the full analytical expression for the $\Pi(q^2)$ function.
We gratefully acknowledge the support of the CERN management and staff and the technical staffs of the participating institutions for their vital contributions. 
This result is part of a project that has received funding from the European Research Council (ERC) under the European Union’s Horizon 2020 research and innovation programme, Grant agreement No. 947715 (POKER). This work was supported by the HISKP, University of Bonn (Germany), JINR (Dubna), MON and RAS (Russia), ETH Zurich and SNSF Grant No. 169133, 186181, 186158, 197346 (Switzerland), and grants FONDECYT 1191103, 1190845, and 3170852, UTFSM PI~M~18~13 and ANID PIA/APOYO AFB180002 and ANID - Millenium Science Initiative Program - ICN2019\_044 (Chile).
\\
The work presented in this paper was done before February 2022.
\end{acknowledgments}
\appendix*
\section{\label{appendixA}Explicit formulas for $\Pi(q^2)$, $\sigma_{\Zpr,Res}$, and $\frac{d\sigma_{\Zpr,Rad}}{dE_F}$}

In this brief appendix, we report the formulas we implemented in \texttt{MADDUMP} in order to correctly account for the $\gamma - \Zpr$ mixing, described in Sec.~\ref{sec:pheno}. This effective mixing, arising from one-loop diagrams involving $\mu$ and $\tau$, results in a $e^\pm - \Zpr$ interaction term $e \Pi(q^2) \Zpr_\mu \left(\overline{e} \gamma^\mu e\right)$; the functional form of $\Pi(q^2)$ is reported in Eq.~\ref{eq:pi}. In order to allow for $\Zpr$ production via electron and positrons, the  $e^\pm - \Zpr$ interaction vertex was added in \texttt{MADDUMP}.
The full $\Pi(q^2)$ formula, computed with \texttt{MATHEMATICA}~\cite{Mathematica}, reads as follows:

\begin{equation}
\begin{split}
\frac{2 \pi^{2}}{e g_{Z^{\prime}}} \Pi\left(q^{2}\right)=
\frac{1}{3}\left[\frac{1}{2} \log \left(\frac{r_{\tau}}{r_{\mu}}\right)+2\left(r_{\mu}-2 r_{\tau}\right)+ \right. \\
 -\left(1+2 r_{\mu}\right) \sqrt{1-4 r_{\mu}} \operatorname{coth}^{-1}\left(\sqrt{1-4 r_{\mu}}\right)+\\
\left.+\left(1+2 r_{\tau}\right) \sqrt{1-4 r_{\tau}} \operatorname{coth}^{-1}\left(\sqrt{1-4 r_{\tau}}\right)\right],\quad
\end{split}
\end{equation}
with $r_{\mu}=m_{\mu}^{2} / q^{2}$ and $r_{\tau}=m_{\tau}^{2} / q^{2}$.

\begin{figure}
    \centering
    \includegraphics[width=.48\textwidth]{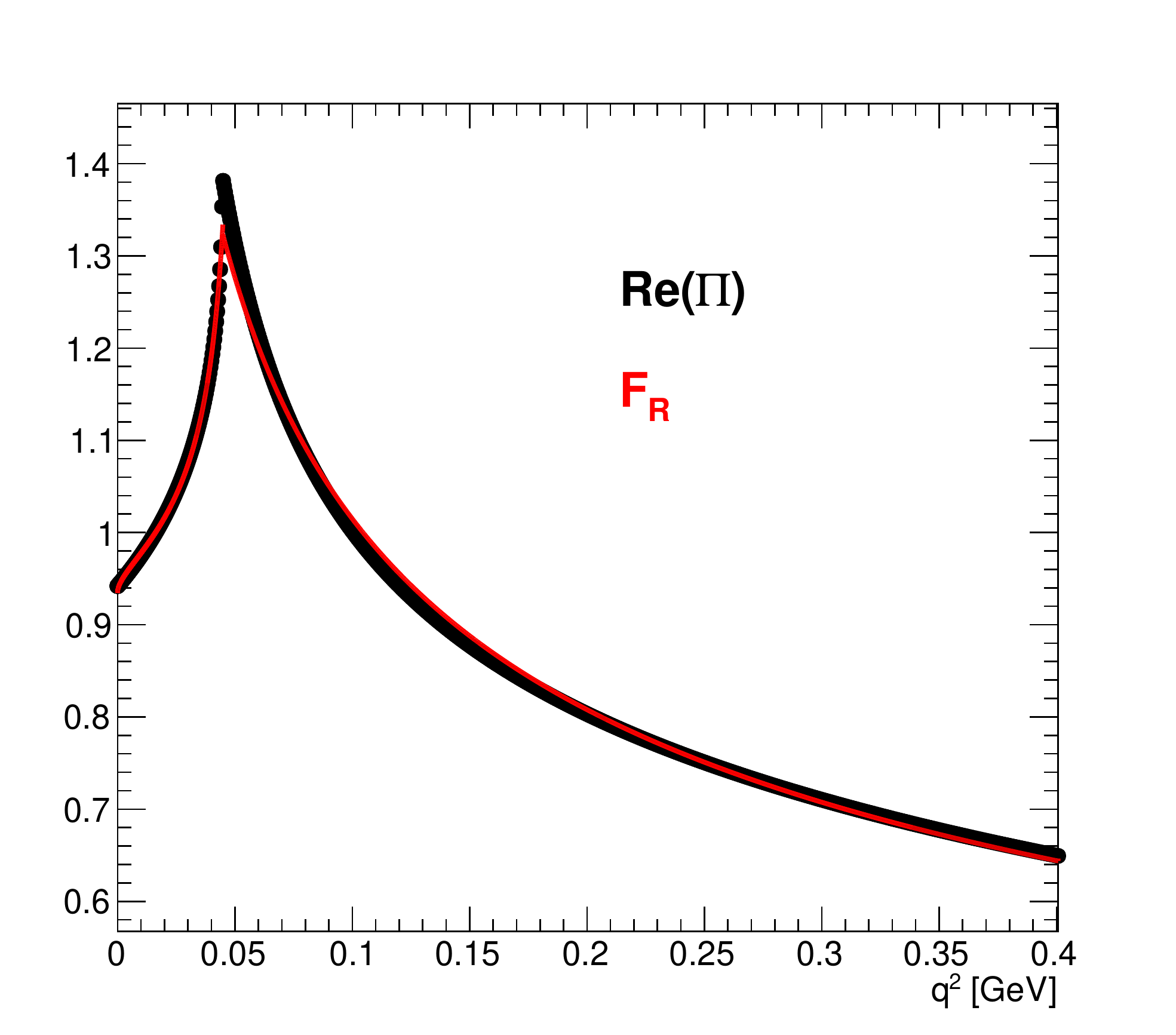}
    \caption{The real part of the $\Pi(q^2)$ function, comparing the numerical result obtained from Eq.~\ref{eq:pi} with the numerical parameterization $F_R$ adopted in this work. For illustration purposes, the arbitrary coupling choice $g_\Zpr=2\pi^2/e$ was made.}
    \label{fig:PiNumerical}
\end{figure}

For simplicity, for the real part of $\Pi(q^2)$ we used an effective parameterization, here denoted by $F_\Re(q^2)$. Fig.~\ref{fig:PiNumerical} shows a comparison between $F_\Re(q^2)$ and $\Re (\Pi(q^2))$, numerically evaluated: the relative error remains below $5\%$ and $\Re (\Pi(q^2)) > F_\Re(q^2)$ over the considered $q^2$ range, resulting in conservative limits on $g^{up}_\Zpr$. 

The imaginary part of $\Pi(q^2)$ can be evaluated analytically, and reads:
\begin{widetext}
\begin{equation}
\Im(\Pi(q^2))=
\left\{
\begin{array}{ll}
\frac{e g_\Zpr  }{12 \pi} \left( 1+ 2\frac{m_{\mu}^2}{q^2} \right) \sqrt{1-4 \frac{m_{\mu}^2}{q^2}}  & \mathrm{if}\; 4m^2_\mu < q^2 < 4m^2_\tau\\
0 & \mathrm{otherwise}
\end{array}
\right.
\end{equation}
\end{widetext}

Similarly, we report below the cross-section formulas for $\Zpr$ production and invisible decay. The total cross section for $\Zpr$ production via $e^+e^-$ annihilation and subsequent decay to a pair of scalar dark sector $\chi$ particles reads:

\begin{equation}
    \sigma_{\Zpr,Res,\chi}=\frac{\pi\alpha_{EM}\alpha_D \left|\Pi(s)\right|^2}{3}
    \frac{s \left(1-4r^2_s\right)^\frac{3}{2}}{(s-m^2_\Zpr)^2+\Gamma_\Zpr^2m^2_\Zpr} \; \;,
\end{equation}

where $s$ is the $e^+$ $e^-$ system invariant mass squared, $r_s\equiv m_\chi/s$, and $\Gamma_\Zpr$ is the total $\Zpr$ width. Similarly, the total cross section for the annihilation process, considering the decay to $\nu_\mu$ or $\nu_\tau$, is:
\begin{equation}
    \sigma_{\Zpr,Res,\nu_\mu+\nu_\tau}=\frac{4\pi\alpha_{EM}\alpha_\Zpr \left|\Pi(s)\right|^2}{3}
    \frac{s}{(s-m^2_\Zpr)^2+\Gamma_\Zpr^2m^2_\Zpr} \; \;.
\end{equation}

Finally, the cross section $\frac{d\sigma_{\Zpr,Rad}}{dE_F}$ for the production of a $\Zpr$ via radiative emission and subsequent invisible decay can be obtained starting from the expression for the emission of an on-shell dark photon $\frac{d\sigma_{\Apr,Rad}}{dE_F}$ (see e.g. Ref.~\cite{Gninenko:2017yus} for the exact tree-level formula), via the relation:
\begin{equation}\label{eq:line-shape}
\begin{split}
\frac{d\sigma_{\Zpr,Rad}}{dE_F dq^2} = \frac{d\sigma_{\Apr,Rad}}{dE_F} \frac{\left| \Pi(q^2) \right|^2}{\varepsilon^2}
\cdot
\\
\cdot
\frac{1}{\pi}\frac{\sqrt{q^2}\,\Gamma_\Zpr B_F}{(q^2-m^2_\Zpr)^2+m^2_\Zpr\Gamma^2_\Zpr} \; 
\end{split} \;\;,
\end{equation}
where $q^2$ is the $\Zpr$ daughter particles invariant mass squared and $B_F$ is the branching fraction for the invisible decay channel.

\bibliographystyle{apsrev4-2}
\bibliography{bibliographyNA64_inspiresFormat.bib,bibliographyNA64exp_inspiresFormat.bib,bibliographyOther_inspiresFormat.bib}

\end{document}